\documentclass[]{article}
\usepackage{lmodern}
\usepackage{amssymb,amsmath}
\usepackage{ifxetex,ifluatex}
\usepackage{fixltx2e} 
\ifnum 0\ifxetex 1\fi\ifluatex 1\fi=0 
  \usepackage[T1]{fontenc}
  \usepackage[utf8]{inputenc}
\else 
  \ifxetex
    \usepackage{mathspec}
  \else
    \usepackage{fontspec}
  \fi
  \defaultfontfeatures{Ligatures=TeX,Scale=MatchLowercase}
\fi
\IfFileExists{upquote.sty}{\usepackage{upquote}}{}
\IfFileExists{microtype.sty}{%
\usepackage{microtype}
\UseMicrotypeSet[protrusion]{basicmath} 
}{}
\usepackage[margin=1in]{geometry}
\usepackage{hyperref}
\hypersetup{unicode=true,
            pdftitle={Rule- and context-based dynamic business process modelling and simulation},
            pdfauthor={Olegas Vasilecas, Diana Kalibatiene and Dejan Lavbič},
            pdfborder={0 0 0},
            breaklinks=true}
\usepackage{natbib}
\usepackage{longtable,booktabs}
\usepackage{graphicx,grffile}
\makeatletter
\def\maxwidth{\ifdim\Gin@nat@width>\linewidth\linewidth\else\Gin@nat@width\fi}
\def\maxheight{\ifdim\Gin@nat@height>\textheight\textheight\else\Gin@nat@height\fi}
\makeatother
\setkeys{Gin}{width=\maxwidth,height=\maxheight,keepaspectratio}
\IfFileExists{parskip.sty}{%
\usepackage{parskip}
}{
\setlength{\parindent}{0pt}
\setlength{\parskip}{6pt plus 2pt minus 1pt}
}
\providecommand{\tightlist}{%
  \setlength{\itemsep}{0pt}\setlength{\parskip}{0pt}}
\ifx\paragraph\undefined\else
\let\oldparagraph\paragraph
\renewcommand{\paragraph}[1]{\oldparagraph{#1}\mbox{}}
\fi
\ifx\subparagraph\undefined\else
\let\oldsubparagraph\subparagraph
\renewcommand{\subparagraph}[1]{\oldsubparagraph{#1}\mbox{}}
\fi

\let\rmarkdownfootnote\footnote%
\def\footnote{\protect\rmarkdownfootnote}

\usepackage{titling}

  \title{Rule- and context-based dynamic business process modelling and
simulation}
    \pretitle{\vspace{\droptitle}\centering\huge}
  \posttitle{\par}
    \author{Olegas Vasilecas, Diana Kalibatiene and Dejan Lavbič}
    \preauthor{\centering\large\emph}
  \postauthor{\par}
    \date{}
    \predate{}\postdate{}

\usepackage{booktabs}
\usepackage{longtable}
\usepackage{array}
\usepackage{multirow}
\usepackage[table]{xcolor}
\usepackage{wrapfig}
\usepackage{float}
\usepackage{colortbl}
\usepackage{pdflscape}
\usepackage{tabu}
\usepackage{threeparttable}
\usepackage{threeparttablex}
\usepackage[normalem]{ulem}
\usepackage{makecell}

\usepackage{amsthm}

\theoremstyle{definition}

\theoremstyle{definition}

\theoremstyle{definition}

\theoremstyle{remark}

\begin{document}
\maketitle

\begin{quote}
Olegas Vasilecas, Diana Kalibatiene and \textbf{Dejan Lavbič}. 2016.
\href{https://doi.org/10.1016/j.jss.2016.08.048}{\textbf{Rule- and
context-based dynamic business process modelling and simulation}},
\href{https://www.sciencedirect.com/journal/journal-of-systems-and-software}{Journal
of Systems and Software \textbf{(JSS)}}, 122, pp.~1 - 15.
\end{quote}

\section*{Abstract}\label{abstract}
\addcontentsline{toc}{section}{Abstract}

The traditional approach used to implement a business process (BP) in
today's information systems (IS) no longer covers the actual needs of
the dynamically changing business. Therefore, a necessity for a new
approach of dynamic business process (DBP) modelling and simulation has
arisen. To date, existing approaches to DBP modelling and simulation
have been incomplete, i.e.~they lack theory or a case study or both.
Furthermore, there is no commonly accepted definition of BDP. Current BP
modelling tools are suitable almost solely for the modelling and
simulation of a static BP that strictly prescribes which activities, and
in which sequence, to execute. Usually, a DBP is not defined strictly at
the beginning of its execution, and it changes under new conditions at
runtime. In our paper, we propose six requirements of DBP and an
approach for rule- and context-based DBP modelling and simulation. The
approach is based on changing BP rules, BP actions and their sequences
at process instance runtime, according to the new business system
context. Based on the proposed approach, a reference architecture and
prototype of a DBP simulation tool were developed. Modelling and
simulation were carried out using this prototype, and the case study
shows correspondence to the needs of dynamically changing business, as
well as possibilities for modelling and simulating DBP.

\section*{Keywords}\label{keywords}
\addcontentsline{toc}{section}{Keywords}

Dynamic business process, Business rules, Context, Simulation, Business
process modelling

\section{Introduction}\label{introduction}

Nowadays business processes (BPs) are dynamic by nature and affected by
a dynamically changing environment. Examples of such dynamic changes of
environment include regulatory adaptations (e.g.~change in raw material
prices), market evolution (e.g.~stock price change), changes in customer
behaviour (e.g.~rapid change in customer needs), process improvement,
policy shifts and exceptions.

Traditional approaches used to model, simulate and implement BP no
longer cover the actual needs of business, which should be more dynamic
in order to be able to compete. There is thus a necessity to conduct
research in the field of DBP modelling, simulation and their automation
in information systems (IS), and to propose new solutions that
correspond to a dynamic business's needs. Moreover, a business needs
tools which support such DBP and allow it to find answers for the
question of what-if.

Consequently, these business needs can be transformed into the
requirements for research as follows:

\begin{itemize}
\tightlist
\item
  proposing a DBP modelling approach,
\item
  proposing a DBP simulation approach,
\item
  proposing an architecture and implementing it into a BP simulation
  engine.
\end{itemize}

Existing approaches to DBP modelling and simulation are incomplete
because they are lacking in theory and in terms of implementation.
Furthermore, there is no commonly accepted definition of DBP. Current BP
modelling and simulation tools are suitable for modelling and simulation
of a static BP that strictly predicts which activities, and in which
sequence, to execute (e.g.~Simprocess or ARIS 9.7). In the best case,
today's tools and proposed methods allow for changing BP by executing
different configurations of the BP, as in \citep{xiao_towards_2011}, or
using templates for modifying BP activity sequences, as in
\citep{eijndhoven_achieving_2008}. This means that the majority of
existing tools and approaches require strict specification of a BP, and
unexpected sequences of BP activities cannot be included during a BP
execution. Therefore, modelling and simulation of a dynamically changing
BP, e.g.~dynamic processes, is a topical, relevant and challenging task.

As advocated in a number of papers, e.g.~in
\citep{mejia_bernal_dynamic_2010, hermosillo_using_2010, milanovic_modeling_2011},
business rules are applicable for ensuring the dynamicity of BP.

The main opportunities of using business rules to ensure dynamicity of
BP lie in the following: each activity in a process is selected
according to the defined conditions at BP runtime; choice of activity
content; and representing the changing DBP context. However, as
presented in the works related to this paper, there is no complete
approach or tool for rule- and context-based DBP modelling and
simulation -- i.e.~none of the analysed tools (IBM Websphere (v.7.0
2014)\footnote{\url{http://www-03.ibm.com/software/products/en/modeler-advanced}.},
Simprocess (v 2015)\footnote{\url{http://simprocess.com}},
Simul8\footnote{\url{http://www.simul8.com}}, AccuProcess\footnote{\url{http://bpmgeek.com/accuprocess-business-process-modeler}}
and ARIS 9.7\footnote{\url{http://www.softwareag.com/corporate/products/new_releases/aris9/more_capabilities/default.asp}}),
which are widely used for BP modelling and simulation, supports changing
business rules during the simulation of DBP. Some approaches, like
\citep{hermosillo_creating_2010}, describe BP dynamicity using BP
pointcuts, where adaptations can be made, or changes of BP are available
in new instances of the BP, but not at the same instance, like in
\citep{xiao_towards_2011}.

In our paper, we propose an approach for rule- and context-based DBP
modelling and simulation. We define DBP as a process with a not-fixed
sequence of activities, i.e.~activities for execution in the BP are
selected according to the rule conditions and changing context of the
environment. Moreover, activities can be changed at the same BP instance
runtime. We propose an approach for changing a sequence of BP
activities, their content, rules for selecting activities in DBP and the
context of a DBP at process runtime. Based on the proposed approach, the
reference architecture and a prototype of a DBP simulation tool was
developed. A case study was carried out using this prototype, and it
shows correspondence between the obtained results and needs of the
dynamically changing business, as well as possibilities for DBP
modelling and simulating.

The balance of this paper is organized as follows. Section
\ref{related-works} presents related works on DBP. Section
\ref{requirements} presents our proposed requirements of rule- and
context-based DBP. Section \ref{approach} describes our approach for
rule- and context-based DBP modelling and simulation and reference
architecture for the rule- and context-based DBP modelling and
simulation system (DRBPSimul). Section \ref{case-study} presents a case
study of an ordering system. Section \ref{results-and-discussion}
presents results and discussion. Finally, Section
\ref{conclusions-and-future-works} concludes the paper.

\section{Related works}\label{related-works}

In our research, we emphasize DBP where content and the sequence of
activities depends on the context of the environment and can be changed
at runtime. In contrast to DBP, static BP has a strict specification,
i.e.~the content and sequence of activities are defined before BP
instance execution and cannot be changed at runtime. The remaining part
of the work discusses four main aspects of a DBP. They are: existing
definitions, models, realizations and simulation of a DBP.

\subsection{Concept of a dynamic business
process}\label{concept-of-a-dynamic-business-process}

However, there is a range of approaches within DBP modelling and
simulation, the difference of these approaches lies in the understanding
of the concept of dynamicity. Therefore, we start our analysis with the
concept of BP dynamicity. The analysis of the related works shows that
four main concepts are popular for defining the ability of a BP to adapt
to the changing environment. They are as follows:

\begin{itemize}
\tightlist
\item
  dynamicity,
\item
  flexibility,
\item
  agility and
\item
  adaptability.
\end{itemize}

\citet{gong_policy_2012} state that the concept of flexibility is more
appropriate in comparison to agility within the context of a BP.
According to them, the concepts of agility and flexibility overlap and
represent a system's ability to respond to changes in the environment.
Agility in this case has a stronger emphasis on the speed aspect than on
flexibility. According to \citet{pucher_agile-_2010}, dynamic BP is a
variant of agile process and enables a business user to make changes in
the process at runtime, for example, by selecting a different
sub-process at predefined decision points. Another kind of a flexible BP
is an ad-hoc BP that has no underlying process definition and consists
of a set of activities relating to possible contents
\citep{pucher_agile-_2010, dustdar_mining_2005, bizagi_understanding_2016}.
In an ad-hoc process, a user decides what to do and when to do it. The
most flexible BP, according to \citet{pucher_agile-_2010}, is adaptive
BP, which is a process that changes according to the environmental
conditions at runtime. Moreover, the execution of adaptive BP instance
influences execution of the next BP instance, e.g.~real-time knowledge
from the last process execution can influence the execution of the next
one.

In spite of the above, in many cases BP dynamicity, flexibility, agility
and adaptability are used as synonyms. For example, organizations like
(Wikipedia\footnote{\url{http://en.wikipedia.org/wiki/Dynamic_business_process_management}},
Gartner, Inc.\footnote{\url{http://www.gartner.com/it-glossary/dynamic-business-process-management-bpm}},
WhatIs.com\footnote{\url{http://whatis.techtarget.com/definition/dynamic-BPM-business-process-management}})
and research papers, like \citep{adams_dynamic_2010} and
\citep{pesic_declarative_2006} define the dynamicity of BP as: the
ability to react to changing conditions (internal and/or external) of
operation, according to the client's individual needs, in an appropriate
and timely manner at process instance runtime without having a negative
impact on the process essence or its expected completion. Management of
such DBP is understood as the ability to support process change by any
role, at any time, with very low latency.

The differences in DBP definitions and modelling and simulation
approaches can be found by analysing the level of dynamicity of BP. As
proposed by WhatIs.com, there are three levels of DBP. The first, and
lowest, level of dynamicity is described as using decision points, in
which a human or an automated system decides what to do next according
to predefined rules. Almost all of today's approaches and tools
implement this level of dynamicity.

The second, and middle, level of dynamicity allows for automatic
configurations of a BP, like choosing an alternative template for
processing activities or changing the order of activities in a process
if conditions change. A number of approaches, proposing dynamic
adaptation using variability models, are presented. They are, like
\citet{hallerbach_issues_2008}, \citet{alferez_dynamic_2014},
\citet{milani_modelling_2016} etc. However, all those methods are based
on the idea that business process consists of variable and not variable
segments and only variable segments of a process could be changed, in
some cases, like in \citep[@][]{alferez_dynamic_2014}, at runtime.

Approaches presented by \citet{eijndhoven_achieving_2008},
\citet{hermosillo_using_2010}, \citet{hermosillo_creating_2010} fall
into this level of dynamicity, since their main idea is to define
decision points in a process and describe those decisions through
business rules. The approach, presented in \citep{xiao_towards_2011},
also falls into this level of dynamicity. Although
\citet{xiao_towards_2011} suggest constructing a process from a number
of reusable fragments, new process schemas generated according to the
new conditions are applicable only at the next process instance
execution \citep{berkane_pattern-based_2012}. In their approach
\citep{hildebrandt_declarative_2011}, the authors propose a Dynamic
Condition Response Graphs (DCRGraphs) model based on an event-based
process model to ensure flexibility and graphical notation of workflows.

The third and the highest level of dynamicity is goal-driven BP that can
be changed over time and at runtime according to the new conditions and
the customer's needs. The approaches, suggested in
\citet{pesic_declarative_2006} and \citet{adams_dynamic_2010}, are the
most refined among those analysed, since they allow for constructing BP
from a number of possible and reusable tasks at runtime. However, the
authors say nothing about the possible goal-orientation of the dynamic
execution of BP.

As can be summarized from the previous analysis, the BP dynamicity can
be viewed through vertical and horizontal dimensions. The vertical
dimension represents the DBP level, which defines the ability of a BP to
adapt to changes in the environment. At the lowest level we have the
least DBP, like, for example, a BP with decision points, and at the
highest level we have the most DBP. This would, for example, be the case
for a BP, which is adaptable at runtime and is goal-driven. The number
of levels of dynamicity of a BP depends on the definition of dynamicity
as a feature. Therefore, in section \ref{requirements} of this paper we
define requirement for DBP. The horizontal dimension can present
implementation perspective, which could be execution, simulation or
automation.

\subsection{Context-sensitive BP}\label{context-sensitive-bp}

As advocated in a number of papers, dynamicity and adaptability of a BP
is ensured through being context-sensitive BP
\citep{saidani_towards_2007, mejia_bernal_dynamic_2010, nunes_dynamic_2011}.
Such a process is able to adapt the execution of its instances to the
changing context \citep{saidani_towards_2007}. Authors of
\citet{laznik_context_2013} uses context for exception handling in BP
execution language. From the BP modelling viewpoint, authors
\citet{rosemann_context-aware_2006} define context as: ``the minimum of
variables containing all relevant information that impact the design and
execution of a BP''. A context-aware modelling framework is introduced
in \citep{coutaz_context_2005}, where the authors define context as a
set of entities, a set of roles that entities may satisfy, a set of
relations between the entities, and a set of situations that denote
specific configurations of entities, roles, or relations. As presented
in the framework proposed by \citet{coutaz_context_2005}, a
context-aware system should include the situation and context
identification layer for identifying changes in situations or in a
context. The authors \citet{mejia_bernal_dynamic_2010} do not specify
how they define, manage and adopt changes in a context.

As can be summarised from the papers analysed above, context can be
internal and external. An internal context is represented by a state of
system resources, and an external context is represented by state of
system environment expressed by a set of variables and rules. When an
external context changes, i.e.~something important happens outside the
system, a system should react to this change. Those external changes,
however, can cause internal changes in a system, and it is also
necessary to react to these changes. For example, an external change in
a context might be a change in share price, to which a system may react
by buying shares. Of course, after this action the amount of money can
be insufficient for other purchases. Therefore, the system should react
to this internal event and, for example, borrow money.

Therefore, two contexts (internal and external) should be taken into
account when modelling dynamic system. And yet, the existing approaches,
like those analysed in
\citep{saidani_towards_2007, mejia_bernal_dynamic_2010, nunes_dynamic_2011},
allow for the definition of one type of context: internal or external
only, but not both. For example, the authors \citet{hu_dynamic_2013}
define context as the current state of system resources. That said, in
some application domains, only one type of context is important.

Other researchers propose to ensure BP dynamicity through an event-based
approach. Authors such as \citet{hermosillo_creating_2010},
\citet{hermosillo_using_2010}, \citet{hermosillo_towards_2012} suggest
using complex event processing to respond dynamically to the actual
state of an environment. However, not all authors define the events,
i.e.~external or internal or both, to which their system reacts. Authors
\citet{chandy_event_2009} propose to extend the BPM engine with an Event
Processing engine, which can obtain information about the events that
happen outside and inside the BP, giving the system complete
context-awareness when making decisions.

The importance of separating external context from internal context lies
in the fact that BP activities could the change internal context of a
system, but not the external context of the system.

\subsection{Rule-based BP}\label{rule-based-bp}

One more direction for DBP modelling is the involvement of business
rules and business rule-based modelling. We believe that the main
opportunities of using business rules to ensure dynamicity of BP are as
follows: each activity in a process is selected according to the defined
conditions at BP runtime, and the content of an activity is chosen based
on the changing internal and/or external context. However, the proposed
rule-based approaches -- some of which are reviewed in this section --
do not cover all these aspects of DBP. In \citep{hermosillo_using_2010}
the authors use rules to define the relationship between complex events
and to filter the interesting ones in their proposed CEVICHE
architecture. When an event that is important for BP adaptation is
detected, the Complex Event Processing (CEP) engine notifies the Aspect
Manager component to adapt the BP (instance or model) with the
corresponding activity (by authors, aspect) at runtime. It is in line
with this that the authors \citet{hermosillo_using_2010},
\citet{bastinos_sasa_sparql-based_2015},
\citet{sasa_ontology-based_2011} concentrate on complex events
processing in their research; they \citep{hermosillo_using_2010} say
nothing about internal and external context description. The authors
\citet{milanovic_modeling_2011} use business rule patterns to enrich BP
by possible cases and to increase BP flexibility. However, their
proposition satisfies only the first level of BP dynamicity according to
WhatIs.com. In \citep{pesic_declarative_2006} the authors use
constraints to define relations among tasks in the ConDec language. At
every moment during the execution of a process model, there is a
judgement about whether or not the model is correct. Authors of
\citep{charfi_aspect-oriented_2004, charfi_reliable_2006} uses the idea
of Aspect-Oriented Programming for combining existing web services into
more sophisticated web services. Authors of \citep{alferez_dynamic_2014}
propose dynamic adaptation of service compositions with variability
models at process runtime, as discussed above.

In \citep{mejia_bernal_dynamic_2010} the authors propose an approach,
which consists of two main steps: decomposing process into ECA
(Event-Condition-Action) rule set and adapting process to the context
data. During the first step the authors describe how to express
transitions between activities in the form of ECA rules, where an event
is generated when an activity has finished its execution, a condition
used to verify which workflow part is enabled, and where a rule action
determines the next activity that has to be executed. During the next
step the adaptation at runtime is performed.
\citet{mejia_bernal_dynamic_2010} use a Control Flow Checking (CFC)
table to represent the dependencies between activities. When some
modification is performed, a new version of the CFC table is generated
and used to update workflow. However, as proposed in
\citep{mejia_bernal_dynamic_2010}, all processes have a strictly defined
sequence of activities and this sequence of activities can be changed
according to the defined set of rules.

\subsection{Implementation aspects of
BP}\label{implementation-aspects-of-bp}

A particular part of the papers analysed is aimed at proposing
approaches for dynamic workflow\footnote{As defined by
  \citet{muehlen_workflow-based_2004}, a workflow is a specific
  representation of a process which is designed in such a way that the
  formal coordination mechanisms between activities, applications, and
  process participants can be controlled by an information system, by
  the so-called workflow management system - i.e.~a workflow is a
  formal, or implementation-specific, representation of a BP.} modelling
and simulation, like in
\citep{traverso_automated_2004, rao_composition_2006, dohring_vbpmn:_2011}.
According to the levels of BP dynamicity proposed by WhatIs.com, almost
all the studies fall into the first, and rarely into the second, level
of dynamicity. The authors \citet{sadiq_managing_2000} discuss different
issues of dynamic workflow modelling according to change and time. They
do not, however, present any approach or method for dynamic workflow
modelling. In \citep{li_dynamic_2012} the authors use product structure
trees for dynamic workflow modelling and visualization, but such an
approach limits the dynamicity of workflows. In
\citep{kwan_dynamic_1997} a framework for dynamic workflow modelling is
presented. Though the framework has four perspectives in which where
behavioural perspective is responsible for dynamic workflow modelling,
the authors say nothing about implementation of the framework or its
detailed application.

In this paper, we have briefly reviewed architectures proposed for
implementing suggested approaches on rule- and context-based DBP
modelling and simulation. However, in \citep{valenca_systematic_2013}
only half of the analysed researches offered automated tool support or
prototypes for DBP modelling and simulation. The authors
\citet{hermosillo_creating_2010}, \citet{hermosillo_using_2010} propose
the CEVICHE framework and deal with architecture for DBP adaptation. As
\citet{pourshahid_systematic_2012} state, the main drawback of the
suggested approach is that the pointcuts and the adaptation rules should
be predefined in the BP model and CEVICHE can only handle known specific
cases. In other words, it is not a generic framework that can be used to
improve the design of the process models. Rather, it executes already
known alternative processes after detecting predefined situations. The
authors \citet{bui_achieving_2013} propose a SOA-based architecture for
their approach implementation. The main components of the architecture -
implementing dynamic workflows - are described as follows. Aspect
manager with Aspect database are responsible for determining advices
from Advice repository, according to calculated values for variables in
general processes. The main functions of Context server are context
management, determining context changes from Adaptors and informing the
Aspect manager about user-context changes. Service Discovery Manager is
responsible for assisting care-receivers by searching for services,
prioritizing them and selecting the most suitable ones. Other components
are described in \citep{bui_achieving_2013}. However, the proposed
architecture is strongly domain-specific, since it was developed for the
healthcare domain. Another solution for the healthcare domain is
presented in \citep{geebelen_mvc_2010}.

\subsection{Analysis of existing tools on BP modelling and
simulation}\label{analysis-of-existing-tools-on-bp-modelling-and-simulation}

The next thing we analyse in this paper is the implementation of the
proposed approaches. The concept of an implementation of the proposed
approach could be viewed from three perspectives: case study,
automation, or simulation. As the analysis shows, the biggest part of
the approaches is implemented as a case study, like in
\citep{mejia_bernal_dynamic_2010, hermosillo_using_2010, kwan_dynamic_1997},
or an automation of the proposed approach at some level. For example, as
presented in \citep{xiao_towards_2011}, the proposed implementation
(i.e.~automation) of the approach allows for changing BP by presenting
several different instances of the changed BP, or by presenting
templates for BP execution, like in \citep{eijndhoven_achieving_2008}.
However, not one of the previously analysed researches presents
simulation of the DBP.

For more comprehensive study, rule-based BP modelling and simulation
tools are compared in Tables \ref{tab:comparison-rule-based-BP-tools}
and \ref{tab:comparison-BRMS}.

\begin{table}

\caption{\label{tab:comparison-rule-based-BP-tools}Comparison of rule-based BP simulation tools ("+" -- has this feature, "+/--" -- has this feature in part, "--" -- does not have this feature).}
\centering
\fontsize{8}{10}\selectfont
\begin{tabu} to \linewidth {>{\raggedright\arraybackslash}p{2cm}>{\raggedright\arraybackslash}p{4cm}>{\centering}X>{\centering}X>{\centering}X>{\centering}X>{\centering}X}
\toprule
Category & Comparison attribute & IBM WebSphere (v.7.0.2014) & Simprocess (v 2015) & Simul8 & Accu Process & ARIS 9.7\\
\midrule
 & Business rule integration into BP & + & + & + & + & +\\
\cmidrule{2-7}
 & Business rule inclusion into BP using decision tables & + & + & + & -- & +\\
\cmidrule{2-7}
 & Model generation using business rule templates & -- & -- & -- & -- & +\\
\cmidrule{2-7}
 & Re-use of business rules and their templates & + & + & -- & -- & --\\
\cmidrule{2-7}
 & Re-use of business rules and their templates in different projects & + & -- & -- & -- & --\\
\cmidrule{2-7}
 & Business rules modelling politics & + & + & -- & -- & --\\
\cmidrule{2-7}
\multirow{-7}{2cm}{\raggedright\arraybackslash Rules usage within BP} & Trigger-based business rule calling & + & + & + & -- & +\\
\cmidrule{1-7}
 & Hierarchical business rule structure & -- & -- & -- & -- & --\\
\cmidrule{2-7}
 & Rules versioning & + & + & + & -- & --\\
\cmidrule{2-7}
\multirow{-3}{2cm}{\raggedright\arraybackslash Rules structuring and versioning} & Changing the initial model data by rules during simulation & + & + & + & + & +\\
\cmidrule{1-7}
 & Error checking in rules & + & + & + & + & +\\
\cmidrule{2-7}
 & Comparison of business rules & + & -- & -- & -- & +/--\\
\cmidrule{2-7}
 & Automatic generation of reports & -- & -- & -- & -- & --\\
\cmidrule{2-7}
\multirow{-4}{2cm}{\raggedright\arraybackslash Ensuring quality of rules} & Analysis of rules after their execution and/or during their execution & + & -- & -- & -- & --\\
\cmidrule{1-7}
 & DBP simulation & +/-- & -- & +/-- & -- & --\\
\cmidrule{2-7}
 & Changing business rules during suspending the simulation of DBP & +/-- & -- & +/-- & -- & --\\
\cmidrule{2-7}
 & Possibility to define a context (internal or external) & -- & -- & -- & -- & --\\
\cmidrule{2-7}
 & Possibility to change context during execution & -- & -- & -- & -- & --\\
\cmidrule{2-7}
\multirow{-5}{2cm}{\raggedright\arraybackslash DBP simulation} & Level of dynamicity according to WhatIs.com & first & first & first & first & first\\
\bottomrule
\end{tabu}
\end{table}

The results of the comparison of rule-based BP modelling tools are
presented in Table 1, where, based on the related works, criteria for
comparison are divided into four groups: business rules usage within BP,
business rules structuring and versioning, ensuring quality of business
rules, and DBP simulation. The summary of the comparison presented in
Table \ref{tab:comparison-rule-based-BP-tools} is as follows:

\begin{itemize}
\item
  \textbf{IBM Websphere Business Modeler Advanced (v.7.0 2014)} is a BP
  modelling and analysis tool. It allows for simulating of dynamic BP by
  choosing an appropriate execution branch. However, it is not suitable
  for changing rules during the suspension of the process instance. It
  is possible to apply changes at the next BP instance.
\item
  \textbf{Simprocess (v 2015)} is a process modelling and simulation
  tool which integrates Process mapping, hierarchical event-driven
  simulation, and activity-based costing (ABC). However, it is not
  suitable for dynamic BP simulation.
\item
  \textbf{Simul8} is a process modelling and simulation tool that allows
  for defining dynamic parameters which can be changed according to the
  defined rules. However, this tool is not suitable for dynamic BP
  simulation.
\item
  \textbf{AccuProcess Modeler} is a visual BP modelling software product
  that helps businesspeople to document, simulate and improve their BP.
  However, it has no dynamic BP simulation functionality.
\item
  \textbf{ARIS 9.7} is a visual BP modelling and simulation software
  product, with incorporated ARIS Business Rules Designer, full support
  of BPMN 2.0 for BP modelling, and static BP simulation included in
  this version of the software.
\end{itemize}

As can be seen from Table \ref{tab:comparison-rule-based-BP-tools},
existing tools are well developed to model, analyse and simulate static
BP. However, those tools are not suitable for simulation of DBP.

Another comparison of nine process management systems (PMS) (YAWL
\citep{hofstede_modern_2010}, DECLARE \citep{aalst_declarative_2009},
OPERA \citep{hagen_exception_2000}, ADEPT2
\citep{goser_next-generation_2007}, ADOME \citep{chiu_logical_2000},
AgentWork \citep{muller_agentwork:_2004}, ProCycle
\citep{weber_providing_2009}, WASA \citep{weske_formal_2001} and SmartPM
+ Continuous Planning) in terms of managing adaptability is presented in
\citep{marrella_continuous_2011}. The authors define adaptability as the
capability to face exceptional changes -- triggered by foreseeable or
unforeseeable events and -- that may require that BP instances be
adapted in order to be carried out. The majority of PMSs use a
case-based reasoning approach\footnote{Case-based reasoning (CBR) is the
  way of solving new problems based on the solutions of similar past
  problems: users are supported to adapt processes by taking into
  account how previously similar events have been managed
  \citep{marrella_continuous_2011}.} to support adaptation of workflow
specifications to changing circumstances. Adaptation remains manual,
since users need to decide how to manage the events by choosing provided
suggestions. Pre-planned approaches to exceptional changes (e.g.~for
each kind of failure that is envisioned to occur, a specific contingency
process is defined a priori) are available in YAWL, DECLARE and OPERA.
An unplanned adaptation is possible in SmartPM + Continuous Planning
only.

To discern additional requirements of DBP and implementation of these
requirements we review existing BRMS. A comparison of the five
well-known BRMSs (Drools\footnote{\url{http://www.drools.org/}}, OpenL
Tablets\footnote{\url{http://openl-tablets.sourceforge.net/}},
OpenRules\footnote{\url{http://openrules.com/}}, IBM ILog BRMS\footnote{\url{http://www-01.ibm.com/software/integration/business-rule-management/jrules-family/}}
and Blaze Advisor\footnote{\url{http://www.fico.com/en/products/fico-blaze-advisor-decision-rules-management-system}})
is presented in Table \ref{tab:comparison-BRMS}. For this comparison the
set of attributes important for BP simulation was chosen (namely,
Dynamic rules, documentation, possibility for use in simulation, open
source, algorithm use, possibility of using several languages, and rules
defined in a high level language).

\begin{table}[!h]

\caption{\label{tab:comparison-BRMS}Comparison of BRMS ("+" -- has this feature, "+/--" -- has this feature in part, "--" -- does not have this feature).}
\centering
\fontsize{8}{10}\selectfont
\begin{tabu} to \linewidth {>{\raggedright\arraybackslash}p{2cm}>{\centering}X>{\centering}X>{\centering}X>{\centering}X>{\centering}X}
\toprule
Comparison attribute & Drools & OpenL Tablets & OpenRules & IBM ILog & Blaze Advisor\\
\midrule
Dynamic rules & -- & -- & +/-- & +/-- & Production rules\\
Documentation & + & +/-- & + & + & +\\
Possibility to use for simulation & + & -- & -- & + & --\\
Open source & + & + & + & -- & --\\
Algorithm used & Direct inference, Rete & Optimized sequential algorithm & Optimized sequential algorithm, Rete & Direct inference & Rules are triggered implicitly\$\textasciicircum{}\{\textbackslash{}text\{A\}\}\$\\
\addlinespace
Possibility of using several languages & -- & -- & + & + & +/--\\
Rules defined in a high level language & +/-- & + & + & + & +\\
\bottomrule
\end{tabu}
\end{table}

\(^{\text{A}}\) The engine monitors the state of the objects and is able
to find out when the conditions of a rule become true, regardless of why
this condition became true \citep{blaze_software_blaze_1999}.

During the process of choosing BRMS an important aspect is the
possibility of defining rules for non-technical business users. A BRMS
should have the feature to define business rules using high-level
language, e.g.~language that is closer to natural speech than
programming language is. Unlike OpenL Tablets and OpenRules, Drools and
IBM ILog BRMS allow users to specify business rules using low-level
(e.g.~close to programming language) languages. Therefore, those BRMS
are more context-adaptable. Nevertheless, no BRMS analysed in Table
\ref{tab:comparison-BRMS} is suitable for dynamic simulation.

\section{Requirements for a rule- and context-based
DBP}\label{requirements}

According to the related works, we propose requirements for a DBP. To
start, we present the definition of the context used in the
requirements:

\begin{enumerate}
\def\labelenumi{\arabic{enumi}.}
\tightlist
\item
  \textbf{Context} refers to internal context and external context.
\item
  An \textbf{external context} is a set of variables and context rules
  (i.e.~business rules which define some policy of the environment)
  defining a particular state of the environment.
\item
  An \textbf{internal context} is a current state of system resources.
\item
  An \textbf{internal context change} is a change to a current state of
  system resources.
\item
  An \textbf{external context change} is a change to a current state of
  environment, like a change of variables describing environment or
  context rules.
\end{enumerate}

The list of our proposed requirements for a DBP is as follows:

\begin{enumerate}
\def\labelenumi{\arabic{enumi}.}
\tightlist
\item
  A DBP should not have a \textbf{predefined sequence of activities}.

  \begin{enumerate}
  \def\labelenumii{\arabic{enumii}.}
  \tightlist
  \item
    Every subsequent activity should be selected according to predefined
    rules and a context. Therefore, every subsequent BP instance may
    differ from the previous instance of the same BP.
  \item
    If there is no activity for further execution at a DBP runtime, it
    should be possible to do the following:

    \begin{enumerate}
    \def\labelenumiii{\arabic{enumiii}.}
    \tightlist
    \item
      to terminate the execution of a DBP instance;
    \item
      to define a new activity and concerning rules for a DBP instance
      execution.
    \end{enumerate}
  \end{enumerate}
\end{enumerate}

Activities in DBP are selected according to the external and internal
context and rules, i.e.~DBP have no predefined sequence of activities.
The advantage of implementing this requirement is that we increase
dynamicity of a process. For example, in e-health process activities are
selected according to the user's context, which describes user's health
state. When a user's context changes, like blood pressure or
temperature, necessary activities, like drug supply, should be selected
for execution immediately. Moreover, in many cases, it is difficult or
impossible to predict all possible changes of a context and predefine
possible sequences of activities. Contrary to DBP, static processes have
predefined sequence of activities and could be adopted only in
predefined places of a process instance, like variation points.

\begin{enumerate}
\def\labelenumi{\arabic{enumi}.}
\setcounter{enumi}{1}
\tightlist
\item
  \textbf{Context-based dynamicity}:

  \begin{enumerate}
  \def\labelenumii{\arabic{enumii}.}
  \tightlist
  \item
    It should be possible to define an external and internal context.
  \item
    A DBP should react to the change in a context.
  \end{enumerate}
\end{enumerate}

As presented in Section \ref{context-sensitive-bp}, external context is
expressed through variables and rules. Internal context describes state
of system resources. A DBP should react to the changes of a context. The
advantage of implementing this requirement is that at process instance
runtime context can change and process should be adopted to those
changes immediately at process instance runtime.

\begin{enumerate}
\def\labelenumi{\arabic{enumi}.}
\setcounter{enumi}{2}
\tightlist
\item
  \textbf{Rule-based dynamicity}:

  \begin{enumerate}
  \def\labelenumii{\arabic{enumii}.}
  \tightlist
  \item
    It should be possible to define new business rules and to change or
    delete existing business rules at BP runtime.
  \item
    A DBP should react to the changes in business rules at process
    instance runtime.
  \item
    Every next activity in a DBP should be selected according to the
    predefined business rules.
  \end{enumerate}
\end{enumerate}

Activities in DBP are selected according to the predefined rules.
However, if it is necessary, additional rules can be defined or existing
rules can be changed at process instance runtime. The advantage of
implementing this requirement is that at instance runtime rules,
expressing constraints of application domain, laws, etc., could change
and, according to this requirement, it should be possible to implement
those changes at process instance runtime.

\begin{enumerate}
\def\labelenumi{\arabic{enumi}.}
\setcounter{enumi}{3}
\tightlist
\item
  \textbf{Any role} involved in DBP execution should \textbf{support DBP
  instance change}, i.e.~change of activities or their sequence,
  according to the context and rules at runtime with possibly low
  latency.
\end{enumerate}

According to this requirement, a DBP implementing system should react to
changes immediately after those changes comes in force. Moreover, any
role, involved in DBP execution, should support any process instance
changes. The advantage of implementing this requirement is that the
change of a context or business rules influences selection of the next
activity immediately after a change comes in force at process instance
runtime. Implementation of this requirement allows rising a process
dynamicity. Otherwise, if we implement changes during next process
instance, a process becomes static. The majority of analysed approaches
(see Sections \ref{related-works} and \ref{results-and-discussion}) do
not meet this requirement.

\begin{enumerate}
\def\labelenumi{\arabic{enumi}.}
\setcounter{enumi}{4}
\tightlist
\item
  Before selecting the next activity, the historical data of instances
  execution of the same DBP should be analysed and \textbf{the selected
  next activity should not cause execution of an unacceptable sequence
  of activities}, regarding to early gained experience.

  \begin{enumerate}
  \def\labelenumii{\arabic{enumii}.}
  \tightlist
  \item
    The historical data of each DBP instance execution should be stored
    in a log file.
  \item
    It should be possible to define executed instances of a DBP as a
    ``good practice'' and a ``bad practice''.
  \item
    It should be possible to select a suitable instance, labelled as a
    ``good practice'', from the historical data for repeated execution.
  \item
    Time, cost, etc. values should be calculated and stored for each
    executed DBP instance.
  \item
    Instances named ``bad instance''" should not be executed.
  \end{enumerate}
\end{enumerate}

This requirement describes that before executing any DBP instance, a
historical data of this DBP should be analysed. The analysis of
historical data allows determining so called ``good'' and ``bad''
instances of the same process. A ``good instance'' means the one that
requires minimum resources, is completed within an optimal to this
process period and the goal of the process is reached. A ``bad
instance'' means that it utilise too many resources for its execution
and/or do not reach a goal. Therefore, such instance is unacceptable.
Moreover, during historical data analysis, it can be found what
conditions, i.e.~state of current context, leads to such ``bad
instances''. This analysis is necessary to prevent execution of ``bad
instances'' and to propose suitable alternative sequence of activities
for execution.

\begin{enumerate}
\def\labelenumi{\arabic{enumi}.}
\setcounter{enumi}{5}
\tightlist
\item
  A DBP execution and selection of the next activity at DBP
  \textbf{runtime should be based on goal-oriented approach}.
\end{enumerate}

In our paper, a goal-oriented approach means that we are focusing on
achieving some state of the system, but not on the way of how to achieve
some state of the system. Therefore, we implement goal through rules,
which together with check, if we approaching to a goal or not, are used
to select each next activity and/or define new activity for execution.
Advantage of implementing this requirement is that we achieve more
dynamicity and strong goal-orientation of BP.

We argue that a BP, which meets all presented requirements, is a DBP.
Implementing all six requirements provides more freedom in BP modelling.
However, this freedom can be constrained by adding business rules, which
depend on the requirements of application domain. Moreover, the
advantage of the implementing the proposed requirements is that it
allows DBP being a goal-oriented.

\section{A rule- and context-based DBP modelling and simulation
approach}\label{approach}

According to the DBP requirements defined in the previous section, we
propose an approach for rule- and context-based DBP modelling and
simulation. The approach is based on the fact that at BP runtime it is
possible to change rules, BP actions and their sequence, depending on
the new business system context.

\begin{figure}

{\centering \includegraphics[width=1\linewidth]{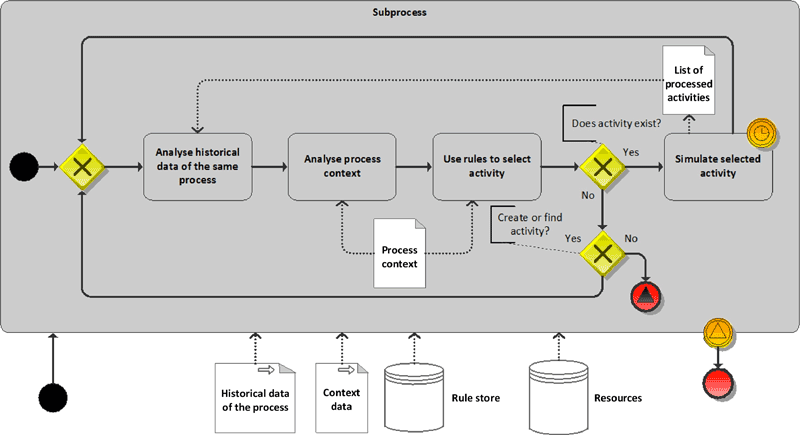}

}

\caption{An approach for rule- and context-based DBP simulation}\label{fig:approach}
\end{figure}

The main idea of the approach is presented in Figure \ref{fig:approach}.
The DBP simulation approach consists of these steps:

\begin{enumerate}
\def\labelenumi{\arabic{enumi}.}
\tightlist
\item
  The historical data of the same BP, e.g.~already executed instances of
  the same BP, are analysed. This activity is useful for distinguishing
  successful and unsuccessful BP simulation instances and using them for
  the current instance simulation.
\item
  The context of the DBP simulation is analysed.
\item
  According to the context, the set of rules are chosen for further
  selection of BP activities.
\item
  According to the rules, an activity is selected for execution. Here we
  may have three cases, as follows:

  \begin{enumerate}
  \def\labelenumii{\arabic{enumii}.}
  \tightlist
  \item
    one activity -- in this case the selected activity is executed and
    we go to the first step;
  \item
    several activities -- in this case the activity with greater
    priority is executed and we go to the first step; and
  \item
    no activities -- in this case the process is finished or we have to
    define a new or select another activity for execution.
  \end{enumerate}
\end{enumerate}

Moreover, an activity uses resources during its execution. For more
about resources see \citep{vasilecas_resource_2015}.

\begin{figure}

{\centering \includegraphics[width=0.8\linewidth]{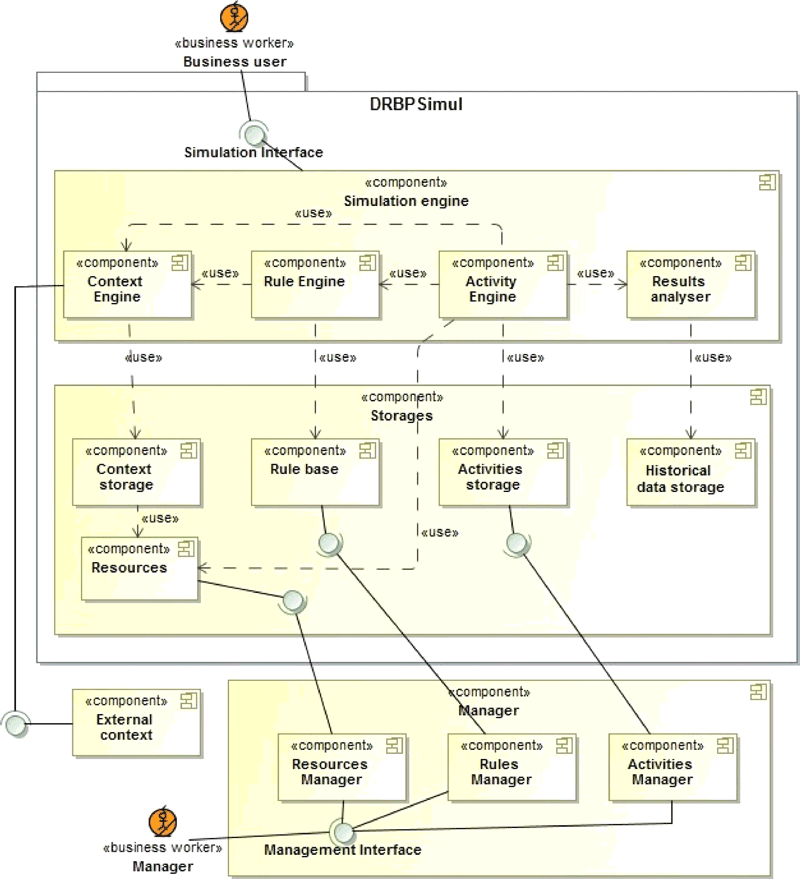}

}

\caption{The architecture of the rule- and context-based DBP modelling and simulation system (DRBPSimul)}\label{fig:architecture}
\end{figure}

In accordance with the proposed rule- and context-based DBP simulation
approach, a system architecture -- DRBPSimul -- was developed. It is
presented in Figure \ref{fig:architecture}.

The proposed architecture (Figure \ref{fig:architecture}) consists of
these components:

\begin{itemize}
\tightlist
\item
  \textbf{Simulation Interface} is responsible for the collaboration of
  the user with the Simulation engine;
\item
  \textbf{Simulation engine} is responsible for the simulation
  execution. It works as described in the model presented in Figure
  \ref{fig:approach}. The Simulation engine consists of these
  components:

  \begin{itemize}
  \tightlist
  \item
    \textbf{The Activity engine}, which is responsible for the
    processing of the selected activities;
  \item
    \textbf{The Rule engine}, which is responsible for the checking of
    the selected rules that correspond to the existing context. The
    Activity engine uses the Rule engine to select the activities for
    execution;
  \item
    \textbf{The Context engine}, which is responsible for the context
    analysis and definition of the process instance context. The Rule
    engine uses the Context engine to determine which rules correspond
    to the existing context and should thus be used for the selection of
    the activities for future execution;
  \item
    \textbf{The Results analyser}, which is used to analyse historical
    data of the executed BP instances. The activity engine uses the
    Results analyser to check if the selected activity for the execution
    causes execution of a not-acceptable sequence of activities.
  \end{itemize}
\item
  \textbf{Storages} (Rule base, Activities storage, Context storage,
  Historical data storage and Resources) are responsible for storage of
  rules, activities, context, historical data and resources used for
  simulation execution and results analysis. Storages handle the
  persistence factors of entities (e.g., rules, activities, context,
  historical data and resources) by providing an interface to create,
  delete, and obtain entities and their templates for new entities by
  reference or query.
\item
  \textbf{Manager} (Resources Manager, Rules Manager and Activities
  Manager) -- responsible for managing (i.e.~creating, deleting and
  getting entities their templates for new entities by reference or
  query) of entities through Management Interface.
\item
  \textbf{External context} is a sensor used by Context storage to
  determine an external context for process instance simulation.
\item
  \textbf{Resources} are resources used for the execution of BP. This
  component is used to determine an internal context state and save it
  in Context storage. Moreover, Resources are used by the Activity
  engine to process selected activities. For more about resources see
  \citep{kalibatiene_implementing_2015, vasilecas_resource_2015}.
\end{itemize}

Before the simulation, Manager (see Figure \ref{fig:architecture})
should load activities, rules and resources into Storages and allocate
resources to activities. When the loading is over, an Analyst or
Business user can start the simulation process. The simulation process
is managed through Simulation Interface. E.g., the simulation process
can be started, stopped, continued and modified. The simulation process
is executed and managed by Simulation engine, which gets commands from
Simulation Interface. When Simulation engine gets a message from the
interface to start a simulation process, it sends a message to Context
Engine and Rule Engine to check existing context and to determine which
rules correspond to it and should be used for the selection of the
activities for future execution. After the corresponding rules are
determined, Simulation engine sends a message to Activity Engine to
choose an activity according to the rules. Moreover, Activity Engine
uses Results Analyser to analyse the Historical data of the same DBP.
When activity is selected, it is executed by Activity Engine. After
executing a selected activity, Activity Engine sends results of an
activity execution to Simulation engine, which in turn repeats the
process from determining current context and selecting next activity for
the execution.

The proposed architecture belongs to the reference level, i.e.~it is
independent of the implementation solutions, which could be used to
implement it. For example, a service-oriented approach (SOA) may be used
for the implementation of the architecture, like in
\citep{hermosillo_towards_2012, marrella_continuous_2011, xiao_towards_2011, berkane_pattern-based_2012};
the authors use a SOA for rules execution. This does not mean that other
ways of implementing DBP are neglected.

\section{A case study of rule- and context-based DBP
simulation}\label{case-study}

The proposed rule- and context-based DBP simulation approach was
realised using Microsoft technologies: Visual Studio 2012\footnote{\url{https://www.visualstudio.com/}},
C\#\footnote{\url{https://code.msdn.microsoft.com/site/search?f\%5B0\%5D.Type=ProgrammingLanguage}\&
  f\%5B0\%5D.Value=C\%23\&f\%5B0\%5D.Text=C\%23} and .NET platform
4.5\footnote{\url{https://www.microsoft.com/en-us/download/details.aspx?id=30653}}.
Windows Presentation Foundation\footnote{\url{https://msdn.microsoft.com/en-us/library/aa970268\%28v=vs.110\%29.aspx}}
components were used to develop the user interface. Microsoft
technologies were used because the university where the prototype of the
system was developed has licenses for those Microsoft products. An
implementation, named DRBPSimul, and an experiment were developed in
Information Systems Research Laboratory\footnote{Information Systems
  Research Laboratory, Institute of Applied Computer Science, Faculty of
  Fundamental Sciences, Vilnius Gediminas Technical University,
  \url{http://fm.vgtu.lt/faculties/departments/institute-of-applied-computer-science/departaments/information-systems-research-laboratory/73139}}.

\begin{figure}

{\centering \includegraphics[width=1\linewidth]{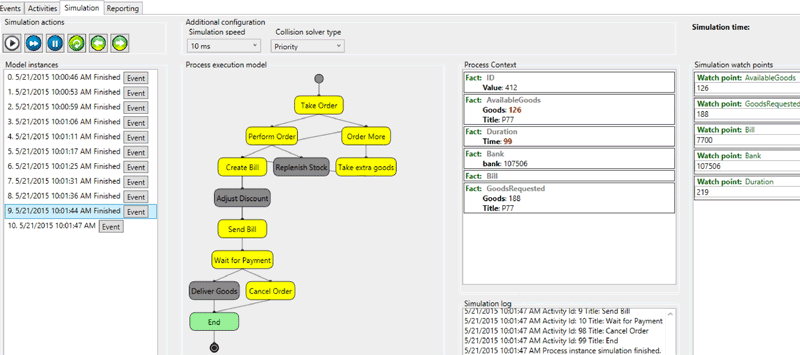}

}

\caption{A simulation example of ordering process}\label{fig:ordering-process}
\end{figure}

An ordering system was used in this case study for demonstrating the
rule- and context-based DBP simulation tool. A simulation example is
presented in Figure \ref{fig:ordering-process}.

The description of Figure \ref{fig:ordering-process} is presented as
follows:

\begin{enumerate}
\def\labelenumi{\arabic{enumi}.}
\item
  The first column on the left presents the list of events, which were
  detected and processed during the simulation. In the presented case,
  the event ``Receive an order'' and its triggered process are presented
  in Figure \ref{fig:ordering-process}.
\item
  The second column on the left presents the graph of the simulated
  process instance. The activities presented in yellow are those
  executed in this process instance simulation; those in grey are
  activities which are not executed in this process instance simulation
  but which were simulated earlier in previous process instances; those
  in green are activities that have just been simulated. As can be seen
  from Figure \ref{fig:ordering-process}, the presentation of a graph
  depends on the executed activities. If all simulated instances are the
  same, we will have a sequential chain of activities. However, if
  simulated instances differ, we will have a graph that is coloured
  differently than the one presented in Figure
  \ref{fig:ordering-process}.
\item
  The third column on the left presents the process context and its
  change during execution of activities. The process context can change
  after execution of each activity. The simulation log is located below
  the process context.
\item
  The fourth column on the left presents simulation ``Watch points''.
\end{enumerate}

\begin{figure}

{\centering \includegraphics[width=1\linewidth]{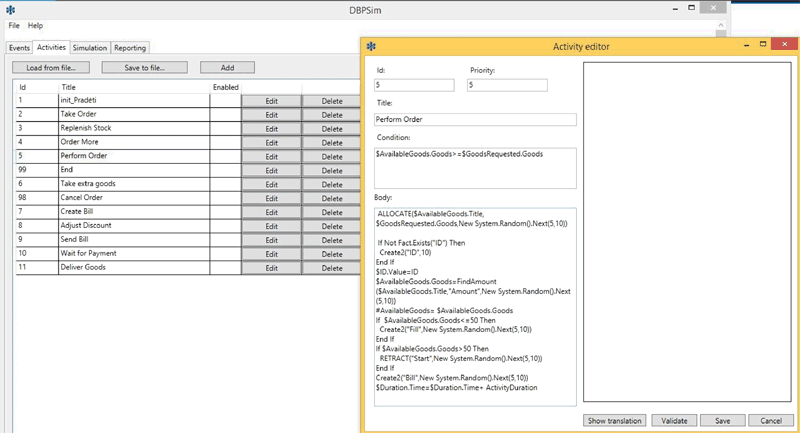}

}

\caption{Example of changing the rules}\label{fig:changing-rules}
\end{figure}

It is possible to change rules and context during the simulation. Rules
could be changed through the Activities tab at the Body placeholder, as
presented in Figure \ref{fig:changing-rules}. After changing a rule, we
can continue a simulation of a BP instance.

The proposed conceptual architecture presented in Figure
\ref{fig:architecture} is now partially realised as follows:

\begin{itemize}
\tightlist
\item
  Simulation Interface, Activity Engine, Rule Engine, Rule base,
  Activities storages and Manager are realised;
\item
  Context Engine, Context Storage and Resources are realised partially;
\item
  Results Analyser and Historical data storage are least realized.
\end{itemize}

\section{Main results and discussion}\label{results-and-discussion}

During our research, we obtained the following results.

The analysis of the related works, including our research conducted on
BP modelling and simulation, allows us to formulate a theoretical
background of a dynamic business process (DBP), as well as to propose an
extended definition and to specify requirements of the rule- and
context-based DBP.

Based on the proposed DBP requirements, an approach for rule- and
context-based DBP modelling and simulation is presented. The main idea
of the approach is that during the simulation of the same BP instance a
rule set and a context can be changed. Those changes can be observed
within the same BP instance immediately after changes come into force,
e.g.~activities and their sequence can be changed at runtime.
Consequently, DBP has no predefined sequence of activities, i.e.~each
subsequent activity for execution is selected after observing an
existing context and rules. Moreover, new activities can be created and
existing activities can be modified at DBP runtime.

To implement the proposed approach, a reference architecture for rule-
and context-based DBP modelling and simulation tool was developed. The
architecture presents the main components and their relationships, which
are necessary to support the proposed approach. Our contribution is
this: unlike in the majority of the analysed research, our proposed
architecture is not domain-specific. Moreover, components and their
relationships are organized in such a way as to support rule- and
context-based DBP modelling and simulation. Additional components, such
as Results analyser linked to Historical data storage, are also included
into the architecture in order to add intelligent functionality to the
approach.

The proposed architecture was implemented into the rule- and
context-based DBP modelling and simulation prototype, and an experiment
on ordering system business processes was carried out. An experiment
shows that instances for the same BP differ in appearance of specific
activities when a different context and rules are applied (see Figure
\ref{fig:ordering-process}). Moreover, we can observe the dynamicity of
a BP.

The comparison of the proposed approach with the other analysed
approaches according to the proposed requirements (see Section
\ref{requirements}) is presented in Table \ref{tab:comparison}. The
first column lists the 11 approaches considered, including our proposed
approach. Each approach is identified by a sequance number and a
reference to the publication/s as follows:

\begin{itemize}
\tightlist
\item
  \textbf{A1} = CEVICHE,
  \citep{hermosillo_creating_2010, hermosillo_using_2010}, where
  implementation includes some examples of code,
\item
  \textbf{A2}, \citep{eijndhoven_achieving_2008}, where implementation
  includes prototype and case study with some examples of code,
\item
  \textbf{A3}, \citep{bui_achieving_2013}, where implementation includes
  prototype with test cases from the homecare domain,
\item
  \textbf{A4}, \citep{yao_modeling_2012}, where implementation includes
  prototype, print screens, from on-boarding customers to IT
  outsourcing,
\item
  \textbf{A5}, \citep{dohring_extended_2010}, where implementation
  includes prototype and print screens,
\item
  \textbf{A6} = PROVOP,
  \citep{hallerbach_managing_2008, hallerbach_issues_2008, hallerbach_context-based_2008, hallerbach_capturing_2010},
  where implementation includes prototype and print screens from
  automotive and healthcare industries,
\item
  \textbf{A7} = DYPROTO,
  \citep{worzberger_dyproto_2011, heer_workflows_2009}, where
  implementation includes prototype, without examples,
\item
  \textbf{A8} = RoDP, \citep{hu_dynamic_2013}, where implementation
  includes example with Mathlab simulation experiment,
\item
  \textbf{A9} = BPFAMA \citep{boukhebouze_rule-based_2011}, where
  implementation includes prototype, print screens and example of
  purchase order,
\item
  \textbf{A10}, \citep{mejia_bernal_dynamic_2010}, where implementation
  includes case study with examples,
\item
  \textbf{A11} = \textbf{Our proposed approach}, where implementation
  includes prototype and print screens.
\end{itemize}

The other columns indicate the requirements and to what extent the given
approach covers each requirement.

\begin{table}[!h]

\caption{\label{tab:comparison}Comparison of the eleven approaches on rule- and context-based DBP modelling and simulation}
\centering
\fontsize{8}{10}\selectfont
\begin{tabu} to \linewidth {>{\raggedright}X>{\centering}X>{\centering}X>{\centering}X>{\centering}X>{\centering}X>{\centering}X>{\centering}X>{\centering}X>{\centering}X>{\centering}X>{\centering}X>{\centering}X>{\centering}X}
\toprule
\multicolumn{1}{c}{Approach} & \multicolumn{1}{c}{R1} & \multicolumn{2}{c}{R2} & \multicolumn{3}{c}{R3} & \multicolumn{1}{c}{R4} & \multicolumn{5}{c}{R5} & \multicolumn{1}{c}{R6} \\
\cmidrule(l{2pt}r{2pt}){1-1} \cmidrule(l{2pt}r{2pt}){2-2} \cmidrule(l{2pt}r{2pt}){3-4} \cmidrule(l{2pt}r{2pt}){5-7} \cmidrule(l{2pt}r{2pt}){8-8} \cmidrule(l{2pt}r{2pt}){9-13} \cmidrule(l{2pt}r{2pt}){14-14}
  &   & R2.1 & R2.2 & R3.1 & R3.2 & R3.3 &   & R5.1 & R5.2 & R5.3 & R5.4 & R5.5 &  \\
\midrule
A1 & No & Yes, external, if defined in CEP. Does not detailed & Yes/No & Yes/No, in CEP & No & No, only when adaptation situation arises & Yes & No & No & No & No & Yes & No\\
A2 & No & No & No & Yes/No & Nothind said & Yes & Yes & No & No & No & No & Yes & No\\
A3 & No & Yes, external$^{\text{B}}$ & Yes & Yes/No & Nothind said & Yes & Yes & Yes & No & No & No & Yes & No\\
A4 & No & Yes, external$^{\text{B}}$ & Yes & Yes/No & Nothind said & Yes & Yes & Yes & No & Yes & No & Yes & Yes\\
A5 & No & Yes$^{\text{C}}$ & Yes & Yes/No & Nothind said & Yes & Yes & No & No & No & No & Yes & No\\
\addlinespace
A6 & No & Yes$^{\text{C}}$ & Yes & Yes/No & Nothind said & Yes & Yes & Yes & Yes & Yes/No & No & Yes & Yes/No\\
A7 & No & No & No & Yes/No & Nothind said & Yes & Yes & No & No & No & No & Yes & No\\
A8 & No & Yes, internal. Does not detailed & Yes & Yes/No & Nothind said & Yes & Yes & No & No & No & No & Yes & No\\
A9 & No & No & No & Yes/No & Nothind said & Yes & Yes & Yes & Yes/No, diagnostic phase & No & No & Yes & No\\
A10 & No & Yes, internal & Yes & Yes/No & Nothind said & Yes & Yes & No & No & No & No & Yes & No\\
A11 & Yes & Yes, both & Yes & Yes & Yes & Yes & Yes & Yes & Yes & Yes & Yes & Yes & Yes\\
\bottomrule
\end{tabu}
\end{table}

\(^{\text{B}}\) The authors \citet{bui_achieving_2013},
\citet{yao_modeling_2012} identify user-context, which can be described
by a set of variables characterizing the user's needs (i.e.~time of
taking medications, blood pressure, etc. in \citep{bui_achieving_2013}).
We make a presumption that it is an external context, since the most
variables belongs on concrete situation (i.e.~a user).

\(^{\text{C}}\) The authors \citet{dohring_extended_2010},
\citet{hallerbach_managing_2008}, \citet{hallerbach_issues_2008},
\citet{hallerbach_context-based_2008}, \citet{hallerbach_capturing_2010}
define context by context variables, and context rules are applied to
define the behaviour of the process in concrete situations. They do not
clearly exclude what type of context (internal or external) is defined.
In most cases, it is the user's needs. Therefore, we make a presumption
that it is an external context, since most variables pertain to a
concrete situation (i.e.~a user).

Reviewing column-by-column, we can observe these results: the analysed
approaches (except for our proposed approach) deal with the initial BP
model, where sequence of process activities can be changed only in
variation points according to the predefined rules.

Regarding the context, the majority of authors describe one type of
context, i.e.~external or internal, since, in the authors' analysed
cases, it suffices to have one type of context - though not all authors
specify how to define context. The common description of a context, also
used by \citet{bui_achieving_2013} and \citet{yao_modeling_2012},
consists of variables and rules.

All the analysed approaches deal with rules, and their application
differs. For example, in
\citep{hermosillo_using_2010, hermosillo_creating_2010} the authors use
rules to define complex events. The authors of the analysed approaches
consider the definition of new rules, or changing or deleting existing
rules, before process simulation (see \textbf{R3.1} in Table
\ref{tab:comparison}). Nothing, however, is said about the possibility
of changing rules at runtime and about how to react to those changes
(see \textbf{R3.2} in Table \ref{tab:comparison}). Moreover, since the
basic ideas of the approaches vary, the analysed approaches differ in
terms of how each subsequent activity is executed in accordance with the
defined rules (see \textbf{R3.3} in Table \ref{tab:comparison}). For
example, in \citep{eijndhoven_achieving_2008} only variable segments of
a process can be changed according to predefined rules. Such a view
limits the degree of dynamicity of a process.

The analysis of the approaches according to process modification at
runtime (see \textbf{R4} in Table \ref{tab:comparison}) shows that all
approaches meet this requirement. However, authors do not show clearly
how to implement this requirement.

As can be seen from Table \ref{tab:comparison}, historical data is used
only in a few approaches, like in \citep{yao_modeling_2012}, PROVOP and
\citep{boukhebouze_rule-based_2011}. But the usage of historical data is
weak. The advantage of selecting ``good practice'' and ``bad practice''
(see \textbf{R5} in Table \ref{tab:comparison}) is not used in the
analysed approaches.

Regarding the \textbf{implementation}, the analysis shows that the
biggest part of the approaches is implemented as examples or case
studies, like in
\citep{mejia_bernal_dynamic_2010, hermosillo_creating_2010, hermosillo_using_2010}
or an automation of the proposed approach at some level, like in
\citep{eijndhoven_achieving_2008}. In half of the analysed studies, the
authors present a small piece of code and some diagrams or examples from
which it is difficult or impossible to determine the level of
implementation of the approaches. Therefore, it is not clear which part
of the method/approach is implemented. Though authors like
\citet{bui_achieving_2013}, \citet{yao_modeling_2012},
\citet{dohring_extended_2010}, PROVOP and
\citet{boukhebouze_rule-based_2011} present prototypes and some print
screens that enhance the proposed approaches, no author presents a
process simulation.

From the presented results, it can be summarized that the main
difference between our proposed approach and those analysed is that ours
is goal-oriented. In such a process, every next activity for execution
is selected according to the context (variables and rules describing
context) and business rules. The analysed approaches deal with an
initial process model, which can be changed at variation points
according to the new user's needs. The main contribution of our proposed
approach is that we ensure dynamicity of a process and allow
implementing goal-oriented approach. Contrary, other approaches are
based on an initial process model that limits the dynamicity and BP can
be changed only in predefined places in a process -- e.g.~variation
points, like in \citep{eijndhoven_achieving_2008, bui_achieving_2013} or
pointcuts, like in
\citep{hermosillo_creating_2010, hermosillo_using_2010}.

Moreover, in our approach, during the process simulation, we can change
rules and react to the changing of context at runtime -- something that
is not supported in all the analysed approaches.

Our proposed approach meets all defined requirements, unlike other
approaches, which meet only a particular set of requirements. This can
be explained by the fact that other studies aim to solve slightly
different types of domain-specific problems -- that is, the majority of
approaches are targeted at meeting the requirements of particular
application domains. In contrast, our approach is general and is not
dependent on a specific domain.

Implementing all six requirements provides more freedom in changing the
BP than is possible in the analysed approaches. However, this freedom
can be constrained by rules. Our developed prototype and obtained
results show that the proposed method is implementable. Moreover, as can
be seen in Figure \ref{fig:ordering-process}, for a BP with the same
goal different sets of activities occur when different rules are applied
when different contexts appear.

Below we try to explain each requirement in term of our presented case
study.

\begin{enumerate}
\def\labelenumi{\arabic{enumi}.}
\item
  Implementation of the first requirement can be seen in Figure
  \ref{fig:ordering-process}. Our process has no predefined sequence of
  activities. Each activity is selected according to the context and
  rules. It can be seen that different instances of the same process has
  different sequence of activities.
\item
  Implementation of the second requirement can be seen in Figure
  \ref{fig:ordering-process} column three (Process Context). In this
  case, we are working with internal context, i.e.~resources.
\item
  Implementation of the third requirement can be seen in Figure
  \ref{fig:changing-rules}, where we present a window with rules, which
  can be changed during paused BP simulation.
\item
  The forth requirement is implemented manually, when it is necessary to
  implement changes, like a rule change, we manually pause the
  simulation, change the rule and continue the simulation of the same
  instance.
\item
  The fifth requirement is implemented by adding rules, which prevent
  execution of ``bad instances''.
\item
  The sixth requirement (goal-orientation) is implemented through
  defined rules, which example can be seen in Figure
  \ref{fig:changing-rules}.
\end{enumerate}

\section{Conclusions and future
works}\label{conclusions-and-future-works}

This analysis of the concept of a dynamic business process (DBP) shows
that there is no commonly accepted DBP definition and neither is there a
complete approach or tool for DBP modelling and simulation. Current BP
modelling and simulation tools are mainly suitable for modelling and
simulating a static BP which strictly predicts activities and the
sequence of their execution. Today's tools and proposed methods allow
us, at best, to change BP by presenting several different instances of
the changed BP or presenting templates for BP execution. This means that
the majority of existing tools and approaches require strict BP
specification of and that an unexpected sequence of BP activities cannot
be executed.

In this paper, we proposed a new approach for rule- and context-based
DBP modelling and simulation. Our main contribution is that we have
proposed the requirements for a DBP and according to those requirements
developed a model for rule- and context-based goal-oriented DBP
modelling and simulation. The main idea of the approach is that during
the simulation of the same BP instance, rules and a context can be
changed and those changes influence selection of the next activity for
execution. Thus, activities and their sequence will be changed within
the same instance immediately after rules and context changes come into
force at runtime. Consequently, the DBP has no strict activities and
their sequence, i.e.~each subsequent activity for execution, is selected
after observing an existing context and rules regarding to the defined
goal.

According to the proposed approach, an architecture for rule- and
context-based DBP modelling and simulation is put forth. This
architecture was implemented into a prototype and the case study shows
correspondence to the needs of the dynamically changing business, as
well as possibilities for modelling and simulating DBP.

The topics for future research are as follows:

\begin{enumerate}
\def\labelenumi{\arabic{enumi}.}
\item
  Refining and improving a set of requirements.

  \begin{enumerate}
  \def\labelenumii{\arabic{enumii}.}
  \tightlist
  \item
    Improving an implementation of a goal-oriented approach in DBP
    simulation.
  \item
    A requirement, that a DBP should meet requirements of a regular BP,
    should be analysed in more details and implemented. This type of
    requirements concerns with correctness and consistency of a DBP.
    Now, this kind of requirements is implemented manually. For example,
    Analyst is responsible for process correctness, since, he defines
    necessary rules and should ensure that deadlocks not arise.
  \end{enumerate}
\item
  Historical data of simulations should be saved in the historical data
  storage and appropriate algorithms should be created and used to
  automate analysis of the stored historical data. During historical
  data analysis, conditions, i.e.~state of current context, which lead
  to ``bad instances'' should be found.
\item
  Improving the developed prototype and making its detailed description.
\item
  Simulating parallel DBP with shared resources.
\end{enumerate}

\end{document}